\documentclass{revtex4}

\usepackage{amsmath,amsfonts,amssymb,amsthm}

\newtheorem{theorem}{\indent Theorem}

\DeclareMathOperator{\A}{A}
\DeclareMathOperator{\B}{B}
\DeclareMathOperator{\AB}{AB}
\DeclareMathOperator{\E}{E}
\DeclareMathOperator{\tr}{Tr}
\DeclareMathOperator{\qu}{qubit}
\DeclareMathOperator{\cb}{cbit}

\DeclareMathOperator{\eb}{ebit}
\DeclareMathOperator{\qus}{qubits}
\DeclareMathOperator{\cbs}{cbits}
\DeclareMathOperator{\ccs}{coherent\ bits}
\DeclareMathOperator{\coh}{cobit}
\DeclareMathOperator{\cohs}{cobits}
\DeclareMathOperator{\ebs}{ebits}

\DeclareMathOperator{\sch}{Sch}
\DeclareMathOperator{\vol}{Vol}

\DeclareMathOperator{\CCE}{CCE}
\DeclareMathOperator{\QQE}{QQE}

\DeclareMathOperator{\QE}{QE}

\def\be{\begin{equation}}
\def\ee{\end{equation}}
\def\bea{\begin{eqnarray}}
\def\eea{\end{eqnarray}}
\def\ben{\begin{eqnarray*}}
\def\een{\end{eqnarray*}}
\def\non{\nonumber}
\def\l{\left}
\def\r{\right}
\def\>{\rangle}
\def\<{\langle}
\def\gab{\gamma_{a' \!, b'}^{a,b}}
\def\gmany{
 |\g_{a \hspace*{-0.2ex} ' \! \oplus x \hs, b \hspace*{-0.1ex} ' \! \oplus y}
    ^{\hspace*{0ex} a \oplus x \hs, \hspace*{0.1ex} b\oplus y}
 \>} 

\def\lbL{ \left[\rule{0pt}{2.4ex}\right. }
\def\rbL{ \left.\rule{0pt}{2.4ex}\right] }

\def\lbm{ \left[\rule{0pt}{2.1ex}\right. }
\def\rbm{ \left.\rule{0pt}{2.1ex}\right] }

\def\lpm{ \left(\rule{0pt}{2.1ex}\right. }
\def\rpm{ \left.\rule{0pt}{2.1ex}\right) }

\def\COCOE{C$\!_{\rm o}\!$C$\!_{\rm o}\!$E}
\def\COQE{C$\!_{\rm o}\!$QE}
\def\QCOE{QC$\!_{\rm o}\!$E}

\newcommand{\smfrac}[2]{\mbox{$\frac{#1}{#2}$}}

\newcommand{\ket}[1]{\mbox{$\left| #1 \right\rangle$}}

\newcommand{\dblbraket}[1]{\mbox{$\langle #1 | #1 \rangle$}}
\newcommand{\eq}[1]{Eq.~(\ref{eq:#1})}

\renewcommand{\sec}[1]{Sec.~\ref{sec:#1}}
\newcommand{\mscite}[1]{Ref.~\cite{#1}}
\newcommand{\mmcite}[1]{Refs.~\cite{#1}}

\newcommand{\upto}[1]{\stackrel{#1}{\approx}}

\def\hs{\hspace*{-0.2ex}}
\def\ot{\otimes}
\def\g{{\gamma}}

\def\e{{\epsilon}}
\def\cG{{\cal G}}
\def\cD{{\cal D}}

\def\cP{{\cal P}}
\def\bbC{\mathbb{C}}
\def\ra{\rightarrow}
\def\la{\leftarrow}
\def\va{\vec{a}}
\def\vb{\vec{b}}
\def\vx{\vec{x}}
\def\pfail{p_{\hspace*{0.1ex} \text{fail}}}

\begin{document}
\title{Bidirectional coherent classical communication}
\author{Aram W.\ Harrow$^1$ and Debbie W.\ Leung$^2$ \vspace*{1ex} }
\address{$^1$MIT Physics Dept., 77 Massachusetts Avenue, Cambridge, 
	MA 02139, USA \vspace*{0.5ex} 
\\
	$^2$MSC 107-81, IQI, Caltech, Pasadena, CA 91125, USA \vspace*{1ex} 
}
\date{\today}
\begin{abstract}

A unitary interaction coupling two parties enables quantum
or classical 
communication in both the forward and backward directions.
Each communication capacity can be thought of as a tradeoff between
the achievable rates of specific types of forward and backward
communication.
Our first result shows that for any bipartite unitary gate,
bidirectional coherent classical communication is no more difficult
than bidirectional classical
communication --- they have the same achievable rate
regions.  Previously this result was known only for the unidirectional
capacities (i.e., the boundaries of the tradeoff).
We then relate the tradeoff for two-way coherent communication to the
tradeoff for two-way quantum communication and the tradeoff for
coherent communication in one direction and quantum communication in
the other.
\end{abstract}

\maketitle

\parskip 1ex
\parindent 0ex


\section{Introduction}
\label{sec:intro}

Quantum communication theory typically studies channels which take an
input quantum system from one party (call her Alice), act on it
possibly with some noise (a trace preserving completely positive
map\cite{NC00bk}) and pass the system onto another party (call him
Bob).  A quantum channel can generate quantum or classical
communication or entanglement at some rate.  The maximum rate at which
each task can be done with arbitrary precision and with an
asymptotically large number of channel uses is called the {\em
capacity}.

A bipartite unitary gate coupling Alice and Bob can achieve
similar tasks, with either party (or both) in the role of sender or
receiver. 
Early studies can be found in
\cite{Prehistory1,Eisert00,Collins00,DVCLP}, focusing on more specific
systems and protocols.  For example, a {\sc cnot} can send a classical
bit from Alice to Bob, or from Bob to Alice or generate one EPR pair.
{\em Asymptotic capacities} of a general bipartite unitary evolution
to communicate and to generate entanglement were formalized in
\mscite{BHLS}.
A general expression for the entanglement capacity was found in
\mmcite{Leifer,BHLS} and that for entanglement-assisted one-way
classical capacity was found in \mscite{BHLS}.
Expressions for various one-way quantum capacities were subsequently
found in \mscite{ccc}, by introducing the concepts of {\em coherent
classical communication} and entanglement recycling.  (Their precise
definitions, as well as concepts throughout the rest of this
paragraph, will be clarified in \sec{setup}).  In particular, \mscite{ccc} 
showed that for any gate, the one-way classical capacity is equal to
its one-way coherent capacity.  This further provides an expression
for the one-way classical capacity assisted by any linear amount of
free entanglement, and allows the one-way quantum capacity and the
{\em remote state preparation capacity} to be expressed in terms of
this one-way classical capacity.

However, the core result for bipartite unitary evolution in
\mscite{ccc}, the equality of the one-way classical capacity and the
coherent capacity, is left open for simultaneous two-way
communication.  
Our main result is a proof of this equality in \sec{bidi-ccc}.
For completeness, we also compare two-way classical
communication and coherent classical communication in the regime
of negative communication rates (i.e., consuming communication to help
produce other resources).
Following similar arguments as in \mscite{ccc}, we list some corollaries.  
These are the two-way remote state preparation capacity  
and quantum capacity in terms of the classical capacity.
Our main result is proved by using a coherent version of a one-time
pad (analogous to that in \mscite{q1tp}).  The reason why a more
direct extension of the proof from \mscite{ccc} fails is given in an
appendix.  A second appendix discusses the implications our results
have on the definition of coherent classical communication.

\section{Framework, definitions, and notations}
\label{sec:setup}

Throughout the paper, we consider communication between two parties,
Alice and Bob.  Systems in their possession are denoted by respective 
subscripts $\A$, $\A_{0,1,\cdots}$ and $\B$, $\B_{0,1,\cdots}$.
System labels are omitted when they are clear from the context.  We
also use superscripts $(\A)$ and $(\B)$ for {\em different} (but
analogous) objects related to Alice and Bob (for example, their
respective local operations).  Exp and log are always base 2.
We will primarily use the trace distance $\frac{1}{2}\|\rho-\sigma
\|_1$ to quantify the proximity of any two states $\rho$ and $\sigma$,
where $\|X\|_1 := \tr\sqrt{X^\dag X}$.
%
%
For two pure states $|\alpha\>, |\beta\>$,
$\smfrac{1}{2} \, \| \, |\alpha\>\<\alpha| {-} |\beta\>\<\beta| \, \|_1 =
\epsilon \; \Leftrightarrow \; |\<\beta|\alpha\>|^2 = 1-\e^2$.
We use $\ket{\alpha} \upto{\epsilon} \ket{\beta}$ as a shorthand for
$\smfrac{1}{2} \, \| \, |\alpha\>\<\alpha| {-} |\beta\>\<\beta| \, \|_1 \leq
\epsilon$.

We now review some definitions and background results, mostly from
\mmcite{BHLS,ccc,DHW-big}.  Let $\{\ket{x}\}_{x=0,1}$ be a basis for
$\bbC^2$.
We first define various resources.  
Let an ebit denote a unit of shared quantum correlation, as quantified
by an EPR pair $\ket{\Phi}_{\AB}=\frac{1}{\sqrt{2}}\sum_{x=0}^1
\ket{x}_{\A}\ket{x}_{\B}$.  Throughout the paper, we omit the
tensor product symbol, $\otimes$, if no confusion may arise.
Following \mscite{ccc}, we denote the ability to communicate a qubit in
the forward direction (from Alice to Bob) as qubit($\ra$), and
mathematically, it corresponds to the isometry $\ket{x}_{\A} \ra
\ket{x}_{\B}$.  Qubit communication in the opposite direction, the
isometry $\ket{x}_{\B} \ra \ket{x}_{\A}$, is denoted qubit($\la$).  
Nonunitary evolution can be viewed as a unitary evolution between all
participating parties, together with an inaccessible one called the
environment denoted by $\E$.  Then, the ability to communicate a
classical bit in the forward direction, denoted as cbit($\ra$), is
given by the linear map $\ket{x}_{\A}\ra\ket{x}_{\B}\ket{x}_{\E}$.  
In contrast, a cobit($\ra$) is given by the map $\ket{x}_{\A}\ra
\ket{x}_{\A}\ket{x}_{\B}$.  A cbit($\la$) and a cobit($\la$) are
defined similarly.
We call cobits {\em coherent} classical communication, and cbits {\em
incoherent} classical communication or simply classical communication.
One can view cobits as cbits in which Alice is given the environment
$\E$ as quantum feedback.  The results of this paper imply that cobits
may be equivalently defined as the ability to send cbits through
unitary means. In Appendix~\ref{app:implications} we will make this
idea precise.

Communication theory is primarily concerned with converting
available resources into desired ones.
Roughly speaking, given two communication resources $X$ and $Y$, we
say that $X\geq rY$ if $X$ can be transformed into $Y$ asymptotically
and approximately at rate $r$, i.e., $\forall \, \delta \, {>} \, 0$,
$\exists N$ such that $\forall \, n \, {\geq} \, N$, $n$ copies
(or uses) of $X$ can be transformed into $\geq n(r-\delta)$ copies (or
uses) of $Y$, in an approximate manner to be defined.
For example, Shannon's noisy coding theorem \cite{shannon} for a
classical channel (i.e.\ a stochastic map) $T$ could be stated as $T
\geq C(T) \cbs$, where $C(T) :=\max_{P(\Xi)} \l[H(\Xi){+}H(T(\Xi)){-}
H(\Xi,T(\Xi))\r]$ is the classical capacity of the channel $T$, $H(\cdot)$
is the entropy of a random variable, and the maximization is over all
distribution $P(\Xi)$ of the input alphabet $\Xi$.
If $X\geq Y$ and $Y\geq X$, then we write that $X=Y$.  For example,
the reverse Shannon theorem \cite{BSST} states that $C(T) \cbs \geq
T$, so that $T_1 = \smfrac{C(T_1)}{C(T_2)} \, T_2$ for any two
classical channels $T_1$, $T_2$ (in the presence of unlimited shared
randomness). 
Another result \cite{ccc} of this type, $2\cohs(\ra) = 1 \eb + 1
\qu(\ra)$, will be used in \sec{bidi-toff} to relate the classical and
quantum capacities of unitary gates.

The definition for $X\geq rY$ is only complete given an error
definition, and a good one should ensure transitivity of resource
inequalities: $X\geq rY$ and $Y \geq sZ$ implies $X \geq rsZ$.
Operationally, the two corresponding resource transformations should
be sufficiently accurate to be composable.  
Mathematically, we say that $X\geq rY$ if there exist {\em vanishing}
sequences of nonnegative numbers, $\{\epsilon_n\}, \{\delta_n\}$, 
and protocols $\cP_n$ each using $X$ at most $n$ times (and other
allowed resources), such that $\cP_n \stackrel{\e_n}{\approx}
Y^{\otimes (r-\delta_n) n}$.  
%
%
Here the notion of approximation $\stackrel{\e_n}{\approx}$ is extended 
from states to operations as 
\bea
	\forall |\psi\> \quad		
	\smfrac{1}{2} \; \| \; {\cal I} \! \ot \! \cP_n (|\psi\>) - 
	      {\cal I} \! \ot \! Y^{\otimes (r-\delta_n) n} (|\psi\>) \; \|_1 
	\leq \e_n 
\,, 
\label{eq:error-def}
\eea
where ${\cal I}$ denotes the identity operation on a {\em reference
system} of dimension given by the input to $\cP_n$.  Including a
reference system in \eq{error-def} ensures that $\cP_n$ and
$Y^{\otimes (r-\delta_n) n}$ transform correlations similarly.  Here,
we use the symbol $Y$ to denote the associated state transformation
enabled by the resource (see \sec{intro} for examples).
We will see examples of what the above means in the next section.

We can now define the achievable classical rate region of a unitary
gate $U$ as the set of points $(C_1,C_2,E)$ such that $U \geq
C_1\cbs(\ra) + C_2\cbs(\la) + E \ebs$.  When $C_1$, $C_2$, or $E$ is
negative, it means that the resource is being consumed; for example,
if $E<0$ and $C_1,C_2\geq 0$, then $U + (-E) \ebs \geq C_1\cbs(\ra) +
C_2\cbs(\la)$ represents entanglement-assisted communication.  This
paper is mostly concerned with $C_1,C_2 \geq 0$ and arbitrary $E$.
Part of the $(C_1,C_2,E)$ achievable region has been characterized,
for the special cases of $C_1, C_2 \leq 0$ (entanglement capacity
\cite{BHLS,Leifer} which is not increased by free classical
communication), $C_2=0, E=-\infty$ (one-way classical communication
with unlimited entanglement assistance \cite{BHLS}, though the actual
protocol requires only finite entanglement assistance) and $C_2=0$
(one-way classical communication with arbitrary entanglement
assistance \cite{ccc}).
We can define the achievable coherent classical rate region of $U$
analogously as the triples $(C_1,C_2,E)$ so that $U \geq
C_1\cohs({\ra}) + C_2\cohs(\la) + E \ebs$.  

Reference \cite{ccc} showed that $U \geq C \cbs(\ra) + E\ebs$ if and
only if $U \geq C \cohs(\ra) + E\ebs$, i.e., the coherent and
incoherent classical rate regions coincide on the planes $C_1=0$ and
$C_2=0$.  
In the next section we prove that the coherent and incoherent rate
regions are identical in the entire $C_1,C_2 \geq 0$ quadrant.  
Other quadrants will be considered for completeness -- this amounts to
understanding how to best use back classical communication.  We will
see that assistance by $\cohs$ only generates entanglement and that
$\cbs$ are useless.
We then apply the result to relate the capacity regions of different
types of forward and backward communication.

\section{Bidirectional coherent classical communication}
\label{sec:bidi-ccc}
\begin{theorem}\label{thm:bidi-ccc}
For any bipartite unitary or isometry $U$ and $C_1,C_2 \geq 0$, 
\bea 
	U & \geqslant & C_1 \cbs(\ra) + C_2 \cbs(\la) + E \ebs 
	\quad {\rm iff}
\label{eq:cbit-toff} 
\\
	U & \geqslant & C_1 \cohs(\ra) + C_2 \cohs(\la) + E \ebs 
\label{eq:cobit-toff} 
\eea
\end{theorem}

{\bf Proof:}~ Since $1\coh \geq 1\cb$, it suffices to prove the
forward implication.  In other words, given the existence of protocols
achieving the resource transformation in \eq{cbit-toff}, we will
construct protocols that achieve the resource transformation in
\eq{cobit-toff}.  We delay the discussion for $E \neq 0$ until
the end of this section.  For now, suppose $E=0$.

$\bullet$ {\em The definition of $\cP_n$}  

Formally, \eq{cbit-toff} indicates the existence of sequences of
nonnegative real numbers $\{\epsilon_n\},\{\delta_n\}$ satisfying
$\epsilon_n, \delta_n {\; \ra \; } 0$ as $n{\; \ra\; }\infty$; a
sequence of protocols $\cP_n = (V_n \! \otimes \! W_n) \, U \, \cdots
\, U \, (V_1 \!  \otimes \! W_1) \, U \, (V_0 \!  \otimes \! W_0)$,
where $V_j,W_j$ are local isometries that may also act on extra local
ancilla systems, and sequences of integers $C_1^{(n)},C_2^{(n)}$
satisfying $nC_1 \geq C_1^{(n)} \geq n(C_1{-}\delta_n)$, $nC_2 \geq
C_2^{(n)} \geq n(C_2{-}\delta_n)$, such that the following success
criterion holds.

Let $a \in \{0,1\}^{C_1^{(n)}}$ and $b \in \{0,1\}^{C_2^{(n)}}$ be the
respective messages of Alice and Bob.  Let $\ket{\varphi_{ab}}:=\cP_n
(\ket{a}_{\A_1}\ket{b}_{\B_1})$.  Note that $\ket{\varphi_{ab}}$
generally occupies a space of larger dimension than $\A_1 \ot \B_1$
since $\cP_n$ may add local ancillas.
To say that $\cP_n$ can transmit classical messages, we require that
local measurements on 
$\ket{\varphi_{ab}}$ can generate messages $b'$ for Alice and $a'$ for
Bob according to a distribution $\Pr(a'b'|ab)$ such that
\be 
\forall_{a,b} \quad
\sum_{a',b'} \smfrac{1}{2}\l| \, 
\Pr(a'b'|ab) - \delta_{a,a'}\delta_{b,b'}\r| \leq \epsilon_n
\label{eq:cc-condition}
\ee
where $a', b'$ are summed over $\{0,1\}^{C_1^{(n)}}$ and
$\{0,1\}^{C_2^{(n)}}$ respectively.  
\eq{cc-condition} follows from applying \eq{error-def} to classical
communication, taking the final state to be the distribution of
the output classical messages.
Since any measurement can be implemented as a joint unitary on the
system and an added ancilla, up to a redefinition of $V_n, W_n$, we
can assume
\be
 \ket{\varphi_{ab}} := \cP_n (\ket{a}_{\A_1}\ket{b}_{\B_1}) 
= \sum_{a'\!,b'} |b'\>_{\A_1} |a'\>_{\B_1} 
|\gamma_{a' \!, b'}^{a,b} \>_{\A_2 \B_2} 
\label{eq:coh-comm} \, \ee
where the dimensions of $\A_1$ and $\B_1$ are interchanged by $\cP_n$,
and $|\gamma_{a'\!,b'}^{a,b}\>$ are subnormalized states with 
$\Pr(a'b'|ab):=\<\gab|\gab\>$ satisfying \eq{cc-condition}.  Thus, for
each $a,b$ most of the weight of $\ket{\varphi_{ab}}$ is contained in
the $|\gamma_{a,b}^{a,b}\>$ term, corresponding to error-free
transmission of the messages.  See Fig.\ I(a).

$\bullet$ {\em The three main ideas for turning classical communication
into coherent classical communication}

We first give an informal overview of the construction and the
intuition behind it.  For simplicity, consider the error-free term 
with $|\gamma_{a,b}^{a,b}\>$ in ${\A_2 \B_2}$.
To see why classical communication via unitary means should be
equivalent to coherent classical communication, consider the special
case when $|\gamma_{a,b}^{a,b}\>_{\A_2 \B_2}$ is independent of $a,b$.  
In this case, copying $a,b$ to local ancilla systems $\A_0,\B_0$
before $\cP_n$ and discarding $\A_2 \B_2$ after $\cP_n$ leaves a state
$\upto{\e_n} \ket{b}_{\A_1} \ket{a}_{\A_0}
\ket{a}_{\B_1}\ket{b}_{\B_0}$---the desired coherent classical
communication. See Fig.\ I(b).
In general $|\gamma_{a,b}^{a,b}\>_{\A_2 \B_2}$ will carry information
about $a,b$, so tracing $\A_2 \B_2$ will break
the coherence of the classical communication.
Moreover, if the Schmidt coefficients of $|\gamma_{a,b}^{a,b}\>_{\A_2
\B_2}$ depend on $a,b$, then knowing $a,b$ is not sufficient to
coherently eliminate $|\gamma_{a,b}^{a,b}\>_{\A_2 \B_2}$ without
some additional communication.  The remainder of our proof is built
around the need to coherently eliminate this ancilla.

Our first strategy is to {\em encrypt} the classical messages $a,b$ by
a shared key, in a manner that preserves coherence (similar to that in
\mscite{q1tp}).  The coherent version of a shared key is a maximally
entangled state.  Thus Alice and Bob (1) again copy their messages to
$\A_0, \B_0$, then (2) encrypt, (3) apply $\cP_n$, and (4) decrypt.
Encrypting the message makes it possible to (5) almost decouple the
message from the combined ``key-and-ancilla'' system, which is
approximately in a state $|\Gamma_{00}\>$ independent of $a,b$ (exact
definitions will follow later).
(6) Tracing out $|{\Gamma}_{00}\>$ gives the desired coherent
communication.  Let $\cP_n'$ denote steps (1)-(5) (see Fig.\ I(c)).



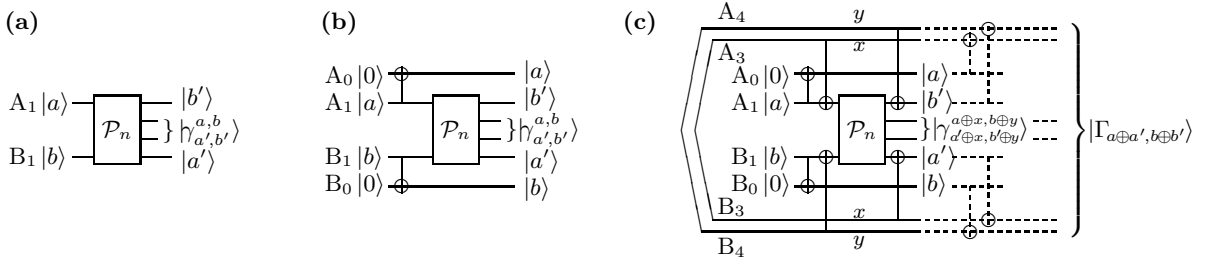
\begin{figure}[h]
\centering \setlength{\unitlength}{0.60mm}
\begin{picture}(245,55)

\put(137,50){\makebox{\bf (c)}}

\put(153,27.5){\line(1,5){4}}
\put(153,27.5){\line(1,-5){4}}
\put(157,47.5){\line(1,0){45}}
\put(157,7.5){\line(1,0){45}}

\put(150,27.5){\line(1,5){4.5}}
\put(150,27.5){\line(1,-5){4.5}}
\put(154.5,50){\line(1,0){47.5}}
\put(154.5,5){\line(1,0){47.5}}

\put(175,21.5){\line(1,0){10}}
\put(175,33.5){\line(1,0){10}}

\put(178,40){\circle{3}}
\put(178,33.5){\line(0,1){8}}
\put(178,21.5){\line(0,-1){8}}
\put(178,15){\circle{3}}

\put(182,33.5){\circle{3}}
\put(182,47.5){\line(0,-1){15.5}}
\put(182,5){\line(0,1){18}}
\put(182,21.5){\circle{3}}

\put(198,33.5){\circle{3}}
\put(198,50){\line(0,-1){18}}
\put(198,7.5){\line(0,1){15.5}}
\put(198,21.5){\circle{3}}

\multiput(202,50)(2,0){16}{{\line(1,0){1}}}
\multiput(202,5)(2,0){16}{{\line(1,0){1}}}
\multiput(202,47.5)(2,0){16}{{\line(1,0){1}}}
\multiput(202,7.5)(2,0){16}{{\line(1,0){1}}}

\multiput(210,40)(2,0){6}{{\line(1,0){1}}}
\multiput(210,15)(2,0){6}{{\line(1,0){1}}}
\multiput(210,21.5)(2,0){6}{{\line(1,0){1}}}
\multiput(210,33.5)(2,0){6}{{\line(1,0){1}}}

\multiput(228,25.5)(2,0){3}{{\line(1,0){1}}}
\multiput(228,29.5)(2,0){3}{{\line(1,0){1}}}

\multiput(214,40)(0,2){5}{{\line(0,1){1}}}
\multiput(218,33.5)(0,2){9}{{\line(0,1){1}}}

\multiput(214,15)(0,-2){6}{{\line(0,-1){1}}}
\multiput(218,21.5)(0,-2){8}{{\line(0,-1){1}}}

\put(218,50){\circle{3}}
\put(214,47.5){\circle{3}}
\put(218,7.5){\circle{3}}
\put(214,5){\circle{3}}

\put(195,21.5){\line(1,0){7}}
\put(195,33.5){\line(1,0){7}}

\put(175,15){\line(1,0){27}}
\put(175,40){\line(1,0){27}}

\put(195,25.5){\line(1,0){7}}
\put(195,29.5){\line(1,0){7}}

\put(201,26.5){\makebox{
$\} | \hs \gamma $}\raisebox{0.3ex}{\tiny 
$
^{a \hs {\oplus} \hs x \hs, b \hs \oplus \hs y}
_{\hs a \! {'} \! \oplus \hs x \hs, b \hs ' \! \hs \oplus \! y}
$}{$\hs \>$}} 


\put(185,20){\framebox(10,15){$\cP_n$}}

\put(158,52){\makebox{$\A_4$}}
\put(158,43){\makebox{$\A_3$}}
\put(161,38){\makebox{$\A_0$}}
\put(161,32){\makebox{$\A_1$}}

\put(161,20){\makebox{$\B_1$}}
\put(161,14){\makebox{$\B_0$}}
\put(158,9){\makebox{$\B_3$}}
\put(158,0){\makebox{$\B_4$}}

\put(168,38){\makebox{$|0\>$}}
\put(168,32){\makebox{$|a\>$}}
\put(168,20){\makebox{$|b\>$}}
\put(168,14){\makebox{$|0\>$}}

\put(203,38.5){\makebox{\small $|a\>$}}
\put(203,32.5){\makebox{\small $|b'\>$}}
\put(203,20){\makebox{\small $|a'\>$}}
\put(203,14){\makebox{\small $|b\>$}}

\put(188,2){\makebox{\footnotesize{$y$}}}
\put(188,8){\makebox{\footnotesize{$x$}}}

\put(188,52){\makebox{\footnotesize{$y$}}}
\put(188,45){\makebox{\footnotesize{$x$}}}

\put(230,2.5){\makebox(10,50){$\left. 
\begin{array}{c} {}\\{~}\\{~}\\{~}\\{~}\\{~}\\{~}\\{~} \end{array} \right\} $}}

\put(240,26){\makebox{$|\Gamma_{\!a \hs \oplus \hs a' \hs, b \hs \oplus 
\hs b' \hs }\>$}}

\put(70,50){\makebox{\bf (b)}}

\put(85,21.5){\line(1,0){10}}
\put(85,33.5){\line(1,0){10}}

\put(88,40){\circle{3}}
\put(88,33.5){\line(0,1){8}}
\put(88,21.5){\line(0,-1){8}}
\put(88,15){\circle{3}}

\put(105,21.5){\line(1,0){8}}
\put(105,33.5){\line(1,0){8}}

\put(85,15){\line(1,0){28}}
\put(85,40){\line(1,0){28}}

\put(95,20){\framebox(10,15){$\cP_n$}}

\put(71,38){\makebox{$\A_0$}}
\put(71,32){\makebox{$\A_1$}}

\put(71,20){\makebox{$\B_1$}}
\put(71,14){\makebox{$\B_0$}}

\put(78,38){\makebox{$|0\>$}}
\put(78,32){\makebox{$|a\>$}}
\put(78,20){\makebox{$|b\>$}}
\put(78,14){\makebox{$|0\>$}}

\put(115,39){\makebox{\small $|a\>$}}
\put(115,33){\makebox{\small $|b'\>$}}
\put(115,19){\makebox{\small $|a'\>$}}
\put(115,13){\makebox{\small $|b\>$}}

\put(105,25.5){\line(1,0){5}}
\put(105,29.5){\line(1,0){5}}
\put(111,26.25){\makebox{\small $\} | \! \gab \>$}}

\put(0,50){\makebox{\bf (a)}}

\put(15,21.5){\line(1,0){5}}
\put(15,33.5){\line(1,0){5}}

\put(30,21.5){\line(1,0){7}}
\put(30,33.5){\line(1,0){7}}

\put(30,25.5){\line(1,0){4}}
\put(30,29.5){\line(1,0){4}}
\put(39,33){\small {$|b'\>$}}
\put(39,19){\small {$|a'\>$}}
\put(35,26){\small {$\} \, |\!\gab \>$}}

\put(20,20){\framebox(10,15){$\cP_n$}}

\put(1,32){\makebox{$\A_1$}}
\put(1,20){\makebox{$\B_1$}}

\put(8,32){\makebox{$|a\>$}}
\put(8,20){\makebox{$|b\>$}}

\end{picture}
\label{fig:3protocols}
\caption{Schematic diagrams for $\cP_n$ and $\cP_n'$.
(a) A given protocol $\cP_n$ for two-way classical communication.  
The output is a superposition (over all $a',b'$) of the depicted states, 
with most of the weight in the $(a',b')=(a,b)$ term.  
The unlabeled output systems in the state $|\gab\>$ are $\A_2,\B_2$.
(b) The same protocol with the inputs copied to local ancillas $\A_0,
\B_0$ before $\cP_n$.  If $|\g_{a,b}^{a,b}\>$ is independent of
$a,b$, two-way coherent classical communication is achieved.
(c) The five steps of $\cP_n'$.  Steps (1)-(4) are shown in solid
lines.  Again, the inputs are copied to local ancillas, but $\cP_n$ is
used on messages encrypted by a coherent one-time-pad (the input
$|a\>_{\A_1}$ is encrypted by the coherent version of the key
$|x\>_{\A_3}$ and the output $|a' \hs \oplus x\>_{\B_1}$ is decrypted by
$|x\>_{\B_3}$; similarly, $|b\>_{\B_1}$ is encrypted by $|y\>_{\B_4}$
and $|b' \hs \oplus y\>_{\A_1}$ decrypted by $|y\>_{\A_4}$.  The
intermediate state is shown in the diagram.  Step (5), shown in dotted
lines, decouples the messages in $\A_{0,1},\B_{0,1}$ from
$\A_{2,3,4},\B_{2,3,4}$, which is in the joint state very close to
$|\Gamma_{00}\>$.
}
\end{figure}


If entanglement were free, then our proof of Theorem
\ref{thm:bidi-ccc} would be finished.  However, we have borrowed
$C_1^{(n)}{+}C_2^{(n)}$ ebits as the encryption key and replaced it
with $|\Gamma_{00}\>$.  Though the entropy of entanglement has not
decreased (by any significant amount), $|\Gamma_{00}\>$ is not
directly usable in subsequent runs of $\cP_n'$.  To address this
problem, we use a second strategy of running $k$ copies of $\cP_n'$ in
parallel and performing entanglement concentration of
$|\Gamma_{00}\>^{\otimes k}$ using the techniques of \cite{BBPS96}.
For sufficiently large $k$, with high probability, we recover most of
the starting ebits.  The regenerated ebits can be used for more
iterations of $\cP_n'^{\otimes k}$ to offset the cost of making the
initial $k \lpm \! C_1^{(n)}{+}C_2^{(n)} \! \rpm$ ebits, without
the need of borrowing from anywhere.

However, a technical problem arises with simple repetition of
$\cP_n'$, which is that errors accumulate.  In particular, a na\"\i ve
application of the triangle inequality gives an error $k \e_n$ but
$k$, $n$ are not independent.  In fact, the entanglement concentration
procedure of \mscite{BBPS96} requires $k\gg
\sch(|\Gamma_{00}\>) = \exp(O(n))$ and we cannot guarantee that $k\epsilon_n
\rightarrow 0$ as $k,n \ra\infty$.  Our third strategy is to treat the
$k$ uses of $\cP_n'$ as $k$ uses of a slightly noisy channel, and
encode only $l$ messages (each having $C_1^{(n)}, C_2^{(n)}$ bits in
the two directions) using classical error correcting codes.  The error
rate then vanishes with a negligible reduction in the communication
rate and now making no assumption about how quickly $\epsilon_n$
approaches zero.  We will see how related errors in decoupling and
entanglement concentration are suppressed.   

We now describe the construction and analyze the error in detail.  

$\bullet$ {\em The definition of $\cP_n'$}
\vspace*{-2ex}
\begin{enumerate}
\item[0.]  
Alice and Bob begin with inputs $\ket{a}_{\A_1}\ket{b}_{\B_1}$ and the
entangled states $\ket{\Phi}^{\! \ot C_1^{(n)}}_{\A_3 \B_3}$ and
$\ket{\Phi}^{\! \ot C_2^{(n)}}_{\A_4 \B_4}$.  (Systems $3$ and $4$
hold the two separate keys for the two messages $a$ and $b$.)  The
initial state can then be written as
\be
\frac{1}{\sqrt{N}}
\sum_x \ket{xx}_{\A_3 \B_3}
\sum_y \ket{yy}_{\A_4 \B_4}
~\ket{a}_{\A_1}\ket{b}_{\B_1} 
\ee
where $x$ and $y$ are summed over $\{0,1\}^{C_1^{(n)}}$ and
$\{0,1\}^{C_2^{(n)}}$, and $N = \exp \lpm \! C_1^{(n)} {+} C_2^{(n)}
\! \rpm$.

\item
They coherently copy the messages to $\A_0, \B_0$.

\item 
They encrypt the messages using the one-time-pad 
$\ket{a}_{\A_1} \ket{x}_{\A_3} \ra \ket{a\oplus x}_{\A_1}\ket{x}_{\A_3}$
and
$\ket{b}_{\B_1} \ket{y}_{\B_4} \ra \ket{b\oplus y}_{\B_1}\ket{y}_{\B_4}$
coherently to obtain
\be
\ket{a}_{\A_0} \ket{b}_{\B_0} \; 
\frac{1}{\sqrt{N}}\sum_{xy}
\ket{x}_{\A_3}\ket{y}_{\A_4} 
\ket{x}_{\B_3}\ket{y}_{\B_4} 
~\ket{a \hs \oplus \hs x}_{\A_1}
\ket{b \hs \oplus \hs y}_{\B_1}
\,.
\ee
\item
Using $U$ $n$ times, they apply $\cP_n$ to registers $\A_1$ and $\B_1$,
obtaining an output state 
\be
\ket{a}_{\A_0} \ket{b}_{\B_0}
\frac{1}{\sqrt{N}}\sum_{xy}
\ket{x}_{\A_3} \ket{y}_{\A_4} 
\ket{x}_{\B_3} \ket{y}_{\B_4} 
\sum_{a',b'} 
| b' \! \oplus y\>_{\A_1} |a' \! \oplus x\>_{\B_1} \gmany_{\A_2 \B_2}
\,.
\label{eq:ready-to-measure}
\ee

\item Alice decrypts her message in $\A_1$ using her key $\A_4$ and Bob
decrypts $\B_1$ using $\B_3$ coherently as 
$|b' \oplus y\>_{\A_1} |y\>_{\A_4} \ra |b'\>_{\A_1} |y\>_{\A_4}$
and
$|a' \oplus x\>_{\B_1} |x\>_{\B_3} \ra |a'\>_{\B_1} |x\>_{\B_3}$
producing a state 
\be
\ket{a}_{\A_0} \ket{b}_{\B_0}
\frac{1}{\sqrt{N}}\sum_{xy}
\ket{x}_{\A_3} \ket{y}_{\A_4} 
 \ket{x}_{\B_3} \ket{y}_{\B_4} 
\sum_{a',b'}
\ket{b'}_{\A_1} \ket{a'}_{\B_1} \gmany_{\A_2 \B_2}
\,. \ee
\item 
Further {\sc cnot}s $\A_1 \ra \A_4$, $\A_0 \ra \A_3$, $\B_1 \ra \B_3$
and $\B_0 \ra \B_4$ will leave $\A_{2,3,4}$ and $\B_{2,3,4}$ almost
decoupled from the classical messages.  To see this, the state has become 
\bea
%
& & \ket{a}_{\A_0}  \ket{b}_{\B_0}
\sum_{a',b'} \ket{b'}_{\A_1}\ket{a'}_{\B_1}
\frac{1}{\sqrt{N}}\sum_{xy}
\ket{a\oplus x}_{\A_3} 
\ket{a'\oplus x}_{\B_3} 
\ket{b'\oplus y}_{\A_4}
\ket{b\oplus y}_{\B_4} 
\gmany_{\A_2 \B_2}
\non \\ & = & 
 \ket{a}_{\A_0}  \ket{b}_{\B_0}
\sum_{a',b'} \ket{b'}_{\A_1}\ket{a'}_{\B_1}
\; \ket{\Gamma_{a\oplus a', b\oplus b'}}_{\A_{2,3,4} \B_{2,3,4}}
\label{eq:phiab}
\,, 
\eea
where 
\bea 
\ket{\Gamma_{\! a\oplus a' \!, b\oplus b' \hs}}_{\A_{2,3,4} \B_{2,3,4}} := 
\frac{1}{\sqrt{N}} \sum_{xy}
\ket{a\oplus x}_{\A_3} 
\ket{a'\oplus x}_{\B_3} 
\ket{b'\oplus y}_{\A_4}
\ket{b\oplus y}_{\B_4} 
\gmany_{\A_2 \B_2} \,.
\eea
The fact $\ket{\Gamma_{\! a\oplus a' \!, b\oplus b' \hs}}$ depends
only on $a \oplus a'$ and $b \oplus b'$, without any other dependence
on $a$ and $b$, can be easily seen by replacing $x,y$ with $a\oplus x,
b\oplus y$ in $\sum_{xy}$ in the RHS of the above.
Note that $\<\Gamma_{\! a\oplus a' \!, b\oplus b' \hs} | \Gamma_{\!
a\oplus a' \!, b\oplus b' \hs} \> = \smfrac{1}{N}\sum_{xy}
\Pr(a'\oplus x, b'\oplus y \, | \, a \oplus x , b \oplus y)$, so in
particular for the state corresponding to the error-free term, we have
$\<\Gamma_{00}|\Gamma_{00}\> = \smfrac{1}{N}\sum_{xy} \Pr(xy|xy) :=
1 - \bar{\e}_n \geq 1-\epsilon_n$ \cite{average}.

Suppose that Alice and Bob could project onto
the space where $a'=a$ and $b'=b$, and tell each other they
have succeeded (by using a little extra communication); then the
resulting ancilla state $\smfrac{1}{\sqrt{1{-}\bar{\e}_n}}
|\Gamma_{00}\>$ has at least $C_1^{(n)} {+} \, C_2^{(n)} {+} \log
(1{-}\e_n)$ ebits, since its largest Schmidt coefficient is 
$\leq \lbm \exp(C_1^{(n)}{+}C_2^{(n)} ) (1{-}\bar{\e}_n) \rbm^{-1/2}$ and
$\bar{\e}_n \leq \e_n$.  (A similar state was studied in \mscite{BHLS}
in the proof that the entanglement capacity of a unitary gate was at
least as large as its classical communication capacity.)  
Furthermore, $|\Gamma_{00}\>$ is manifestly independent of $a,b$.
We will see how to improve the probability of successful projection
onto the error free subspace by using block codes for error
correction, and how correct copies of $|\Gamma_{00}\>$ can be
identified if Alice and Bob can exchange a small amount of
information.
\end{enumerate}

$\bullet$ {\em Main idea on how to perform error correction}

As discussed before, $|\Gamma_{00}\>$ cannot be used directly as an
encryption key -- our use of entanglement in $\cP_n'$ is not
catalytic.
Entanglement concentration of many copies of $|\Gamma_{00}\>$ obtained
from many runs of $\cP_n'$ will make the entanglement overhead for the
one-time-pad negligible, but errors will accumulate.
The idea is to suppress the errors in many uses of $\cP_n'$ by error
correction.
This has to be done with care, since we need to simultaneously ensure
low enough error rates in both the classical message and the state to
be concentrated, as well as sufficient decoupling of the classical
messages from other systems.

Our error-corrected scheme will have $k$ parallel uses of $\cP_n'$,
but the $k$ inputs are chosen to be a valid codeword of an error
correcting code.  Furthermore, for each use of $\cP_n'$, the state in
$\A_{2,3,4} \B_{2,3,4}$ will only be collected for entanglement
concentration if the error syndrome is trivial for that use of
$\cP_n'$.  We use the fact that errors occur rarely (at a rate of
$\epsilon_n$, which goes to zero as $n\ra\infty$) to show that (1) most
states are still used for concentration, and (2) communicating the
indices of the states with non trivial error syndrome requires a
negligible amount of communication.


$\bullet$ {\em Definition of $\cP_{nk}''$: error corrected version of
$(\cP_n')^{\ot k}$ with entanglement concentration}

We construct two codes, one used by Alice to signal to Bob and one
from Bob to Alice.  We consider high distance codes.  The distance of
a code is the minimum Hamming distance between any two codewords,
i.e.\ the number of positions in which they are different.

First consider the code used by Alice.  Let $N_1=2^{C_1^{(n)}}$.
Alice is coding for a channel that takes input symbols from
$[N_1]:=\{1,\ldots,N_1\}$ and has probability $\leq \epsilon_n$ of
error on any input (the error rate depends on both $a$ and $b$).
We would like to encode $[N_1]^l$ in $[N_1]^k$ using a code with
distance $2k\alpha_n$, where $\alpha_n$ is a parameter that will be
chosen later.  Such a code can correct up to any $\lfloor k\alpha_n
{-} \smfrac{1}{2} \rfloor$ errors (without causing much problem, we
just say that the code corrects $k\alpha_n$ errors).  Using standard
arguments \cite{goodcode}, we can construct such a code with $l \geq k
\lbm 1 {-} 2\alpha_n {-} H_2(2\alpha_n)/C_1^{(n)} \rbm $, where
$H_2(p)=-p\log p-(1{-}p)\log (1{-}p)$ is the binary entropy.
The code used by Bob is chosen similarly, with $N_2=2^{C_2^{(n)}}$
input symbols to each use of $\cP_n'$.  For simplicity, Alice's and
Bob's codes share the same values of $l$, $k$ and $\alpha_n$.   
We choose $\alpha_n \geq \max(1/C_1^{(n)}, 1/C_2^{(n)})$ so that $l
\geq k(1{-}3\alpha_n)$.  

Furthermore, we want the probability of having $\geq k\alpha_n$ errors to be
vanishingly small.  This probability is $\leq\exp(-k
D(\alpha_n\|\epsilon_n)) \leq \exp(k + k\alpha_n\log\epsilon_n)$
(using arguments from \cite{CT}) $\leq \exp(-k)$ if
$\alpha_n \geq -2/\log\epsilon_n$.  

Using these codes, Alice and Bob construct $\cP_{nk}''$ as follows
(with steps 1-3 performed coherently).
\vspace*{-2ex} 
\begin{enumerate} 
\item[0.]  
Let $(a^{\rm o}_1, \cdots, a^{\rm o}_l)$ be a vector of $l$ messages
each of $C_1^{(n)}$ bits, and $(b^{\rm o}_1, \cdots, b^{\rm o}_l)$ be
$l$ messages each of $C_2^{(n)}$ bits.
\item 
Using her error correcting code, Alice encodes $(a^{\rm o}_1, \cdots,
a^{\rm o}_l)$ in a valid codeword $\vec{a} = (a_1, \cdots, a_k)$ which
is a $k$-vector.  Similarly, Bob generates a valid codeword $\vec{b} =
(b_1, \cdots, b_k)$ using his code.  
\item  
Let $\vec \A_1 := \A_1^{\otimes k}$ denote a tensor product of $k$ input
spaces each of $C_1^{(n)}$ qubits.  Similarly, $\vec \B_1:=
\B_1^{\otimes k}$.  (We will also denote $k$ copies of $\A_{0,2,3,4}$,
and $\B_{0,2,3,4}$ by adding the vector symbol.)
Alice and Bob apply $(\cP_n')^{\otimes k}$ to $\ket{\vec{a}}_{\vec
\A_1} |\vec{b}\>_{\vec \B_1}$; that is, in parallel, they apply
$\cP_n'$ to each pair of inputs $(a_j, b_j)$.  The resulting state is
a tensor product of states of the form given by \eq{phiab}:
\be
\bigotimes_{j=1}^k \lbL
\ket{a_j}_{\A_0}  \ket{b_j}_{\B_0}
\sum_{a_j',b_j'} |b_j'\>_{\A_1} |a_j'\>_{\B_1}
\; |\Gamma_{a_j \oplus a_j', b_j \oplus b_j'}\>_{\A_{2,3,4} \B_{2,3,4}} 
\rbL .
\label{eq:Pn-parallel}
\ee

Define $|\Gamma_{\va \oplus \va', \vb \oplus \vb'}\>_{\vec\A_{234}
\vec\B_{234}} := \bigotimes_{j=1}^k |\Gamma_{a_j \oplus a_j', b_j
\oplus b_j'}\>_{\A_{2,3,4} \B_{2,3,4}}$.  Then, \eq{Pn-parallel} can be
written more succinctly as
\be 
\ket{\va}_{\vec \A_0} |\vb\>_{\vec \B_0} 
 \sum_{\va',\vb'} |\vb'\>_{\vec \A_1} \ket{\va'}_{\vec \B_1}
|\Gamma_{\va \oplus \va', \vb \oplus \vb'}\>_{\vec\A_{234} \vec\B_{234}} \,.
\ee

\item
Alice performs the error correction step on $\vec \A_1$ and Bob
does the same on $\vec \B_1$.  According to our code
constructions, this (joint) step fails with probability $\pfail\leq
2\cdot 2^{-k}$.  (We will see below why $\pfail$ is
independent of $\va$ and $\vb$.)

In order to describe the residual state, we now introduce $\cG_{\hs
\A} = \{\vx \, {\in} \, [N_1]^k : |\vx| \, {\leq} \, k\alpha_n\}$ and
$\cG_{\B} = \{\vx \,{\in}\, [N_2]^k : |\vx| \, {\leq} \, k\alpha_n\}$,
where $|\vx|:=|\{j : x_j \,{\neq}\, 0\}|$ denotes the Hamming weight
of $\vx$.  Thus $\cG_{\hs \A,\B}$ are sets of correctable (good)
errors, in the sense that there exist local decoding isometries
$\cD_{\hs \A},\cD_{\B}$ such that for any code word $\va\in [N_1]^k$
we have $\forall \va'\in \va\oplus\cG_{\hs \A}, \cD_{\hs \A}\ket{\va'}
= \ket{\va}\ket{\va\oplus \va'}$ (and similarly, if $\vb\in[N_2]^k$ is
a codeword, then $\forall \vb'\in \vb\oplus\cG_{\B}, \cD_{\B} |\vb'\>
= |\vb\> |\vb\oplus \vb'\>$).  For concreteness, let the decoding maps
take $\vec \A_1$ to $\vec \A_1 \vec \A_5$ and $\vec \B_1$ to $\vec
\B_1 \vec \B_5$.

Conditioned on success, Alice and Bob are left with
\bea
& & \frac{1}{\sqrt{1{-}\pfail}} \, |\va,\vb\>_{\vec \A_{0,1}} 
		|\va,\vb\>_{\vec \B_{0,1}} 
\sum_{\va' \hs \in \va \oplus \cG_{\hs \A}} 
\sum_{~\vb' \hs \in \vb \oplus \cG_{\B}}
|\vb \oplus \vb'\>_{\vec \A_5} |\va \oplus \va'\>_{\vec \B_5}
|\Gamma_{\va\oplus \va', \vb\oplus \vb'}\>_{\vec \A_{234} \vec \B_{234}}
\\ &:=& \frac{1}{\sqrt{1{-}\pfail}} \, |\va,\vb\>_{\vec \A_{0,1}} 
		|\va,\vb\>_{\vec \B_{0,1}} 
 \sum_{\va'' \hs \in\cG_{\hs \A}} \sum_{~\vb'' \hs \in \cG_{\B}}
|\vb''\>_{\vec \A_5} |\va''\>_{\vec \B_5}
|\Gamma_{\va'',\vb''}\>_{\vec \A_{234}\vec \B_{234}},
\eea
where we have defined $\va'' := \va\oplus \va'$ and $\vb'' :=
\vb\oplus\vb'$.  Note that $2^{-k+1} \geq \pfail =
\sum_{(\va'',\vb'')\not\in \cG_{\hs \A} \times \cG_{\B}}
\dblbraket{\Gamma_{\va'',\vb''}}$, which is manifestly independent
of $\vec a,\vec b$.
The ancilla is now {\em completely} decoupled from the message,
resulting in coherent classical communication.  The only remaining
issue is recovering entanglement from the ancilla, so for the
remainder of the protocol we ignore the now decoupled states $|\vec a,
\vec b\>_{\vec\A_{0,1}} |\vec a,\vec b\>_{\vec\B_{0,1}}$.


\item
For any $\vx$, define $S(\vx):=\{j : x_j \,{\neq}\, 0\}$ to be set of
positions where $\vx$ is nonzero.  If $\vx\in\cG_{\hs \A}$ (or
$\cG_{\B}$), then $|S(\vx)| \leq k\alpha_n$.  Thus, $S(\vx)$ can be
written using $\leq \log \sum_{j \leq k\alpha_n} \!\! \binom{k}{j}
\leq \log \binom{k}{k \alpha_n} 
+ \log (k\alpha_n) \leq kH_2(\alpha_n) + \log (k \alpha_n)$ bits.

The next step is for Alice to compute $|S(\vb'')\>$ from $|\vb''\>$
and communicate it to Bob using $\lpm \! kH_2(\alpha_n) + \log (k\alpha_n) 
\! \rpm \cbs(\ra)$.  Similarly, Bob sends $\ket{S(\va'')}$ to Alice using
$\lpm \! kH_2(\alpha_n) + \log (k\alpha_n) \! \rpm \cbs(\la)$.  Here
we need to assume that some (possibly inefficient) protocol to send
$O(k)$ bits in either direction with error $\exp(-\,{k}\hs-\hs{1})$ (chosen for
convenience) and with $Rk$ uses of $U$ for some constant $R$.  Such a
protocol was shown in \mscite{BHLS} and the bound on the error can be
obtained from the HSW theorem \cite{HSW}.

Alice and Bob now have the state
\be \frac{1}{\sqrt{1{-}\pfail}} \,
\sum_{\va''\in\cG_{\hs \A}} \sum_{\vb''\in\cG_{\B}}
|S(\va'') S(\vb'')\>_{\vec\A_6} \, |\vb''\>_{\vec\A_5} \, 
|S(\va'') S(\vb'')\>_{\vec\B_6} \, |\va''\>_{\vec\B_5} \,  
|\Gamma_{\va'',\vb''}\>_{\vec\A_{234} \vec \B_{234}}.
\ee
Conditioning on their knowledge of $S(\va''),S(\vb'')$, Alice and Bob
can now identify $k'\geq k(1-2\alpha_n)$ positions where
$a_j''=b_j''=0$, and extract $k'$ copies of
$\smfrac{1}{\sqrt{1{-}\pfail}}|\Gamma_{00}\>$.
Note that leaking $S(\va''), S(\vb'')$ to the environment will not
affect the extraction procedure, therefore, coherent computation and
communication of $S(\va''), S(\vb'')$ is unnecessary.  (We have not
explicitly included the environment's copy of $|S(\va'') S(\vb'')\>$ 
in the equations to minimize clutter.) 
After extracting $k'$ copies of
$\smfrac{1}{\sqrt{1{-}\pfail}}|\Gamma_{00}\>$, we can safely discard
the remainder of the state, which is now completely decoupled from both
$\lbm \!  \smfrac{1}{\sqrt{1{-}\pfail}}|\Gamma_{00}\> \! \rbm
^{\otimes k'}$ and the message $|\vec a\>_{\A_0} |\vec b\>_{\A_1}
|\vec b\>_{\B_0} |\vec a\>_{\B_1}$.

\item
Alice and Bob perform entanglement concentration ${\cal E}_{\rm conc}$
(using the techniques of \mscite{BBPS96}) on
$\lbm \! \smfrac{1}{\sqrt{1{-}\pfail}}|\Gamma_{00}\> \! \rbm^{\otimes k'}$.  
Note that since $\smfrac{1}{\sqrt{1{-}\pfail}}|\Gamma_{00}\>$ can be
created using $U$ $n$ times and then using classical communication and
postselection, it must have Schmidt rank $\leq {\rm Sch}(U)^n$, where
${\rm Sch}(U)$ is the Schmidt number of the gate $U$ \cite{Nielsen98}.
Also recall that $E \lbm \! \smfrac{1}{\sqrt{1{-}\pfail}}|\Gamma_{00}\> 
\!\rbm \geq C_1^{(n)} + C_2^{(n)} + \log (1{-}\e_n)$.
According to \mscite{BBPS96}, ${\cal E}_{\rm conc}$ requires no
communication and with probability 
$\geq 1 - \exp \lbm{-}{\rm Sch}(U)^n \lpm\! \sqrt{k'}-\log(k'{+}1) \!\rpm 
\rbm$
produces at least $k' \lbm C_1^{(n)} {+} C_2^{(n)} {+} \log (1{-}\e_n) \rbm 
- {\rm Sch}(U)^n \lbm \sqrt{k'} {-} \log(k'{+}1) \rbm$ ebits.

\end{enumerate}

$\bullet$ {\em Error and resource accounting} 

$\cP_{nk}''$ consumes a total of \\[1ex]
$~~~$(0) $nk$ uses of $U$ (in the $k$ executions of $\cP_n'$) \\
$~~~$(1) $Rk$ uses of $U$
	(for communicating nontrivial syndrome locations) \\
$~~~$(2) $k \lbm \!\! C_1^{(n)} {+} C_2^{(n)} \!\! \rbm $ ebits 
	(for the encryption of classical messages). \\[1ex]
$\cP_{nk}''$ produces, with probability and fidelity no less than 
$1 \, {-} \, 2 \, {\cdot} \, 2^{{-}(k-1)}
-\exp \lbm \!\!  {-}{\rm Sch}(U)^n \lpm \!\!
\sqrt{k'} {-} \log(k'{+}1) \!\!\rpm \!\!\rbm $, at least \\[1ex]
$~~~$(1) $l \, C_1^{(n)}{\rm cobits}(\ra) +l \, C_2^{(n)}{\rm cobits}(\la)$  
\\
$~~~$(2) $k' \lpm \! C_1^{(n)}{+}C_2^{(n)} {+} \log (1{-}\e_n)\! \rpm 
- {\rm Sch}(U)^n \lpm \! \sqrt{k'}{-}\log(k'{+}1) \! \rpm$ ebits. 

We restate the constraints on the above parameters: 
$\epsilon_n, \delta_n {\; \ra \; } 0$ as $n{\; \ra\; }\infty$;
$C_1^{(n)} \geq n(C_1{-}\delta_n)$, $C_2^{(n)} \geq
n(C_2{-}\delta_n)$;
$\alpha_n \geq \max(1/C_1^{(n)}, 1/C_2^{(n)}, -2/\log\epsilon_n)$; 
$k' \geq k(1{-}2\alpha_n)$; $l \geq k(1{-}3\alpha_n)$. 
 
We define ``error'' to include both infidelity and the probability of
failure.  To leading orders of $k,n$, this is equal to $2^{{-}(k-2)} +
\exp \lbm \!\!  {-} \sqrt{k} \; {\rm Sch}(U)^n \! \! \rbm $.  We
define ``inefficiency'' to include extra uses of $U$, net consumption 
of entanglement, and the amount by which
the coherent classical communication rates fall short of the classical
capacities.  To leading order of $k,n$, these are respectively
$Rk$, $2\alpha_nk(C_1^{(n)}{+}C_2^{(n)}) + \sqrt{k} \, {\rm Sch}(U)^n \approx 
 2\alpha_nk n(C_1{+}C_2) + \sqrt{k} \, {\rm Sch}(U)^n$,
and $nk(C_1{+}C_2) - l(C_1^{(n)} {+} C_2^{(n)}) \leq
nk(3\alpha_n(C_1 {+} C_2) + 2\delta_n)$.  We would like the error to
vanish, as well as the fractional inefficiency, defined as the
inefficiency divided by $kn$, the number of uses of $U$.
Equivalently, we can define $f(k,n)$ to be the {\em sum} of the error
and the fractional inefficiency, and require that $f(k,n)\ra 0$ as
$nk\ra \infty$.
By the above arguments,
\be f(k,n) \leq 2^{{-}(k-2)} + \exp(-\sqrt{k} \; {\rm Sch}(U)^n)
 + 2\alpha_n(C_1 {+} C_2) + \smfrac{1}{n \sqrt{k}} \; {\rm Sch}(U)^n
 + \frac{R}{n}
 + 3\alpha_n(C_1 {+} C_2) + 2 \delta_n \,.
\label{eq:frac-errors}\ee
Note that for any fixed value of $n$, $\lim_{k\ra\infty} f(k,n) =
5\alpha_n(C_1{+}C_2)+2\delta_n + R/n$.  (This requires $k$
to be sufficiently large and also $k \gg {\rm Sch}(U)^{2n}$.)  Now,
allowing $n$ to grow, we have
\be 
	\lim_{n\ra\infty}\lim_{k\ra\infty} f(k,n) = 0. 
\label{eq:goooal}
\ee 
The order of limits in this equation is crucial due to the dependence 
of $k$ on $n$. 

The only remaining problem is our catalytic use of 
$O(nk)$ ebits.  In order to construct a protocol that uses only $U$,
we need to first use $U$ $O(nk)$ times to generate the starting
entanglement.  Then we repeat $\cP_n''$ $m$ times, reusing the same
entanglement.  The catalyst results in an additional fractional
inefficiency of $c/m$ (for some constant $c$ depending only on $U$)
and the errors and 
inefficiencies of $\cP_n''$ 
add up to no more than $mf(k,n)$.  Choosing $m = \lfloor
1/\sqrt{f(k,n)} \rfloor$ will cause all of these errors and
inefficiencies to simultaneously vanish.  More generally,
\be
\lim_{m\ra\infty}\lim_{n\ra\infty}\lim_{k\ra\infty} \; mf(k,n) +
\frac{c}{m} \; = \; 0 \,.
\ee 
This proves the resource inequality
\be U \geq C_1\cohs(\ra) + C_2\cohs(\la).\ee

$\bullet$ {\em The $E<0$ and $E>0$ cases} 

If $E<0$ then entanglement is consumed in $\cP_n$, so there exists a
sequence of integers $E^{(n)} \leq n(E+\delta_n)$ such that
\be \cP_n \! \left( \ket{a}_{\A_1}\ket{b}_{\B_1}
\ket{\Phi}^{E^{(n)}}_{\A_5 \B_5} \right)
= 
\sum_{a', b'} |b'\>_{\A_1} |a'\>_{\B_1} |\g_{a',b'}^{a,b}\>_{\A_2 \B_2}\,.
\ee
In this case, the analysis for $E^{(n)}=0$ goes through, only with
additional entanglement consumed.  Almost all equations are the same,
except now the Schmidt rank for $|\Gamma_{00}\>$ is upper-bounded by
$\l[\sch(U)2^{E+\delta_n}\r]^n$ instead of $\sch(U)^n$.  In
particular, previous arguments still give \eq{goooal} from the
modified \eq{frac-errors}.

If instead $E>0$, entanglement is created, so for some
$E^{(n)}\geq n(E-\delta_n)$ we have
\be \cP_n \! \left( \ket{a}_{\A_1}\ket{b}_{\B_1} \right)
= 
\sum_{a', b'} |b'\>_{\A_1} |a'\>_{\B_1} |\g_{a',b'}^{a,b}\>_{\A_2 \B_2}\,.
\ee
for $E(|\g_{a,b}^{a,b}\>_{\A_2 \B_2}) \geq E^{(n)}$. 
Again, the previous construction and analysis go through, with an
extra $E^{(n)}$ ebits of entanglement of entropy in $|\Gamma_{00}\>$,
and thus an extra fractional efficiency of $\leq 2\alpha_nE$ in
\eq{frac-errors}.   The Schmidt rank of $|\Gamma_{00}\>$ is still
upper bounded by Sch$(U)^n$ in this case. \hfill \qed

\vspace*{1ex} 
So far, we have focused on the $C_1,C_2 \geq 0$ quadrant.  The
following theorem will relate the achievable regions for coherent and
incoherent classical communication when $C_1 \leq 0$ or $C_2 \leq 0$.

\begin{theorem}\label{thm:bidi-nnn}
For any bipartite unitary or isometry $U$ and $C_1,C_2 \geq 0$, 
\bea 
	C_2 \cbs(\la) + U & \geqslant & C_1 \cbs(\ra) + E \ebs 
	\quad \quad {\rm iff}
\label{eq:1locc} 
\\
	U & \geqslant & C_1 \cbs(\ra) + E \ebs 
	\quad \quad {\rm iff}
\label{eq:celo}
\\
	U & \geqslant & C_1 \cohs(\ra) + E \ebs 
	\quad \quad {\rm iff}
\label{eq:ccelo}
\\
	C_2 \cohs(\la) + U & \geqslant & C_1 \cohs(\ra) + (E{+}C_2) \ebs
\label{eq:1loccc} 
\eea
and
\bea 
	C_1 \cbs(\ra) + C_2 \cbs(\la) + U & \geqslant & E \ebs 
	\quad {\rm iff}
\label{eq:2locc} 
\\
	U & \geqslant & E \ebs 
	\quad {\rm iff}
\label{eq:elo}
\\
	C_1 \cohs(\ra) + C_2 \cohs(\la) + U & \geqslant & (E{+}C_1{+}C_2) \ebs 
\label{eq:2loccc} 
\eea
\end{theorem}

In essence, the rates of unidirectional classical communication with
arbitrary amount of entanglement assistance (or generation) are not
increased by (in)coherent classical communication in the opposite
direction, except for a trivial gain of entanglement when the
assisting classical communication is coherent.

{\bf Proof:}~ 
Using superdense coding to send $2 \cohs$ and supplying the required
$1$ qubit of quantum communication by teleportation (using $2 \cbs + 1
\eb$), we have
\be
	1 \cb + 1 \eb \geqslant 1 \coh \,.
\label{eq:tp-sd}
\ee
The above resource transformation is exact and does not require large
blocks.  Thus, composing it with other protocols poses no extra
problem.

For the first part of the theorem, \eq{1locc} $\Rightarrow$ \eq{celo}
follows from how \mscite{ccc} characterizes the set of $(C_1,\hs E)$
that satisfies \eq{1locc}.  Although the proof in \mscite{ccc} did not
mention back communication, it can be easily modified to show that
free classical communication from Bob to Alice does not change the
capacity.  In essence, the optimal tradeoff curve between $C_1$ and
$E$ has an upper bound that remains valid in the presence of back
classical communication, and the same bound is achieved by a protocol
that uses no back classical communication.  A complete proof of this
fact will also appear in \mscite{aramthesis}.
 
\mscite{ccc} also proved that \eq{celo} $\Leftrightarrow$ \eq{ccelo},
and it is trivial that \eq{ccelo} $\Rightarrow$ \eq{1loccc}.  Finally,
\eq{1loccc} $\Rightarrow$ \eq{1locc} because of \eq{tp-sd}.

For the second part of the theorem, 
\mscite{BHLS} proved that \eq{2locc} $\Rightarrow$ \eq{elo}.
It is trivial that \eq{elo} $\Rightarrow$ \eq{2loccc}.  
Using \eq{tp-sd}, \eq{2loccc} $\Rightarrow$ \eq{2locc}.  

\section{Achievable regions for bidirectional communication}  
\label{sec:bidi-toff}

Bipartite unitary gates can be used for several inequivalent purposes
simultaneously, including some (possibly different) forms of forward
and backward communications and entanglement generation.  It is thus
natural to define their capacities in terms of achievable rate regions
(in $3$-dimensional space) and trade-off surfaces.

For example, let CCE be the achievable rate region $\{(C_1,C_2,E): U
\geqslant C_1 \cbs(\ra) + C_2\cbs(\la) +E \ebs\}$, and \COCOE~be the
achievable rate region $\{(C_1,C_2,E): U \geqslant C_1 \cohs(\ra) +
C_2\cohs(\la) + E \ebs\}$.  Theorems \ref{thm:bidi-ccc} and
\ref{thm:bidi-nnn} provide a mapping between CCC and \COCOE$\;$:
\be
\begin{array}{ccc}
  (C_1,C_2,E) \in \CCE & \Longleftrightarrow & 
		(C_1,C_2,E{-}\min(C_1,0){-}\min(C_2,0)) \in 
	\mbox{\COCOE} \,. 
\label{eq:thm12}
\end{array}
\ee
Finding relations between different capacity regions will simplify our
study of capacities of bipartite unitary gates and elicit their
nonlocal properties.

As a second example of relation of achievable regions, consider remote
state preparation, which is the ability to prepare a quantum state
$|\psi\>$ in the laboratory of the receiver, assuming that the sender
has a classical description of $|\psi\>$ (assuming pure states for
simplicity).  We claim that the achievable region RRE for two-way (but
independent forward and backward) remote state preparation is the same
as $\CCE$.  To prove this, first note that $\infty \cbs \geq \! n \,  
\operatorname{remote\ qubits} \geq n \cb$, where $n 
\operatorname{remote\ qubits}$ denotes the ability to remotely prepare
an $n$-qubit state.  Combining this with the fact that even unlimited
back-communication does not improve classical capacity implies that
$\operatorname{RRE}\subset \CCE$.  On the other hand, \mscite{ccc}
showed that $n\ccs \geq n\operatorname{remote\ qubits}$.  Thus the
first quadrants ($C_1,C_2\geq 0$) of RRE and \COCOE$\;$(and thus $\CCE$)
are the same, and the other quadrants of RRE are related to \COCOE$\;$the
same way that $\CCE$ is: backwards cobits can be used to generate
entanglement, but free backwards remote qubits do not improve the
forward capacity.  This means that $\operatorname{RRE}=\CCE$.

Similarly, define QQE to be the region $\{(Q_1,Q_2,E): U \geqslant Q_1
\qus(\ra) + Q_2\qus(\la) +E \ebs\}$, corresponding to two-way quantum
communication.  We can also consider coherent classical communication
in one direction and quantum communication in the other; let \QCOE~be
the region $\{(Q_1,C_2,E) : U \geqslant Q_1 \qus(\ra) + C_2\cohs(\la)
+E \ebs\}$ and define \COQE~similarly.

Ref.~\cite{ccc} related the one-way tradeoff curves 
C$\!_{\rm o}\!$E and QE, defined as 
${\rm C}\!_{\rm o}\!{\rm E}=\{(C,E) : (C,0,E)\in \mbox{\COCOE} \}$ and 
$\QE=\{(Q,E) : (Q,0,E)\in \QQE\}$.  There it was claimed that 
\be
 (Q,E)\in\QE \Leftrightarrow (2Q, E-Q)\in {\rm C}\!_{\rm o}\!{\rm E} \,.
\label{eq:one-way-toff}
\ee
We now rephrase the proof of \eq{one-way-toff} in a form that readily
extends to a relation between entire achievable rate regions (for
different types of bidirectional communication).
\eq{one-way-toff} is due to the equivalence $2\cohs = 1\qu + 1\eb$.
Note that this equivalence involves resource transformations that are
exact and do not require large blocks.  Thus, composing these
transformations with other protocols poses no extra problem, and the
equivalence can be used ``freely.''
To prove \eq{one-way-toff}, choose any $(Q,E)\in\QE$.  
Then $U \geq Q \qus + E\ebs = 2Q \cohs + (E-Q)\ebs$, so 
$(2Q, E-Q)\in {\rm C}\!_{\rm o}\!{\rm E}$.  Conversely, if
$(2Q,E-Q)\in {\rm C}\!_{\rm o}\!{\rm E}$, then 
$U\geq 2Q \cohs + (E-Q)\ebs = Q \qus + E \ebs$, so $(Q,E)\in \QE$.

Note that the above argument still works if we replace $U$ with a
different resource, such as $U-Q_2\qus(\la)$.  Therefore, the same
argument that proved \eq{one-way-toff}
also establishes the following equivalences for bidirectional rate
regions:
\be
\begin{array}{ccc}
(Q_1,Q_2,E)\in\QQE & \Longleftrightarrow & (2Q_1, Q_2, E-Q_1) \in\mbox{\COQE} 
\\[2ex] \Updownarrow & & \Updownarrow
\\[2ex] (Q_1,2Q_2, E-Q_2) \in \mbox{\QCOE} & \Longleftrightarrow & 
	(2Q_1,2Q_2,E-Q_1-Q_2)\in \mbox{\COCOE}
\end{array} .
\ee
Finally, \eq{thm12} further relates QQE, QCE, CQE, CCE, where QCE and
CQE are defined similarly to \QCOE~and \COQE~but with incoherent
classical communication instead.

Thus once one of the capacity regions (say \COCOE) is determined, all 
other capacity regions discussed above are determined.  


\appendix
\section{Why we cannot use the techniques in \mscite{ccc}}

In this appendix, we review the proof of Prop.\ 1 in \mscite{ccc} (the
unidirectional communication analogue of Theorem 1) and show how it
breaks down when applied to two-way communication.  

We first review HSW coding \cite{HSW}, since the proof of Prop.~1 in
\cite{ccc} is based on it.  Given a channel which maps a classical
input $i$ to a quantum state $\rho_i$, the HSW theorem states that its
classical capacity is $C := \max_p S(\sum_i p_i\rho_i) - \sum_i
p_iS(\rho_i)$, where the maximization is over probability
distributions $p$ and $S(\rho):=-\tr\rho\log\rho$ is the von Neumann
entropy.  The HSW theorem can be proved by random coding followed by
expurgation.  That is, we choose $2^{n(C-\delta_n)}$ length $n$
codewords according to the product distribution $p^n(i_1,\ldots, i_n)
= p(i_1)\cdots p(i_n)$ (with $\delta_n \ra 0$ as $n \ra \infty$). Then
with high probability the codewords will on average be almost
perfectly distinguishable from one another.  We then discard (or
``expurgate'') the worst half of the codewords in order to signal with
asymptotically vanishing maximum error at a rate approaching $C$.

Instead of choosing codewords according to $p^n$, we could instead
randomly choose typical sequences (meaning that the frequency of a
letter $i$ is $np_i\pm O(\sqrt{n})$).  In fact, since there are only
$\text{poly}(n)$ different type classes, we can choose all our
codewords to be the same type and still achieve capacity $C$
asymptotically.  (The ``type'' of a string denotes the number of times
each letter appears in the string.)

Now we review the application of the HSW theorem to coherent
communication in Prop.~1 of \cite{ccc}.  Given a gate $U$ such that
$U\geq C \cbs(\ra)$, we know (similar to \eq{coh-comm}) that there
exists a sequence of unitary protocols $\cP_n$, each can communicate a
bit string of length $\approx n (C-\delta_n)$ bits up to an error of
$\epsilon_n$ for $\delta_n \ra 0$, $\e_n \ra 0$.  
$\cP_n$ can be viewed as a channel with HSW capacity $\approx nC$,
i.e., by HSW coding, $\cP_n$ can be used $k$ times, sending $\approx
nkC$ bits with {\em overall} error rate vanishing as $k \ra \infty$.
(This idea was used in \cite{BS02} to bound the size of the ancilla
systems used in unitary gate communication.)

Let $p$ be the distribution that almost achieve the HSW capacity.  Let
$\vec{a} = (a_1, \cdots, a_k)$ be any HSW codeword.  Running $\cP_n$
$k$ times produces the state
$|\varphi\> = \bigotimes_{i=1}^k \cP_n \ket{a_i}_{\A_1}$.  
Alice could have copied the input before the protocol, and by the
construction of the HSW code, Bob can extract $\vec{a}$ with
negligible error and disturbance to $|\varphi\>$, and Alice and Bob
will have possession of a state which is $k \e_n$ close to 
$|\vec{a}\>_{\A_0} 
 |\vec{a}\>_{\B_1} 
 \bigotimes_{i=1}^k (\cP_n \ket{a_i})_{\A_2 \B_2}$. 
The state $|\vec{a}\>$ in $\A_0$ and $\B_1$ will allow Alice and Bob
to coherently reorder the $k$ copies of $\cP_n \ket{a_i}$ (with
preagreed total order of the set of all $nC$-bit words).
The reordered state has no information on $\vec{a}$ except for the
letter frequency.  
Thus, when all $\vec{a} = (a_1, \cdots, a_k)$ are of the same type,
the reordered state becomes independent of $\vec{a}$ and can be
discarded without breaking coherence of the communication of
$\ket{\vec{a}}$.
Or when all $\vec{a}$ are typical sequences, the small information on
$\vec{a}$ can be removed with $O(\sqrt{k})$ qubits of communication.
Here, $k$ and $n$ are independent, so that indeed $k \e_n \ra 0$. 

(The original form of the HSW theorem in which we simply choose random
codewords according to $p^n$ and expurgate causes a problem in this
application.  With high probability, the codewords are typical, but
some codewords can be highly nontypical, with corresponding ancilla
that cannot be made identical to a ``typical ancilla'' using
negligible resources.) 

The same-type HSW coding technique cannot be easily applied in the
two-way case.  Even if Alice only uses HSW codewords $\ket{\vec{a}}$
of the same type and similarly for codewords $|\vec{b}\>$ of Bob, the
joint string $(\vec{a},\vec{b}):=((a_1,b_1),\ldots (a_k,b_k))$ need
not have the same type.  With high probability $(\vec{a},\vec{b})$
will be typical, but some are far from typical.  Worst still, these
are composite codewords that depend {\em jointly} on $\vec{a}$ and
$\vec{b}$ and cannot be expurgated by independent expurgation of
individual codewords used by Alice and Bob.

Thus we obtain the strange situation where the average error is small,
but we cannot make the maximum error small because expurgation
requires a linear amount of communication.  A similar problem was
found in bidirectional classical channels, where the achievable
capacity regions are different depending on whether average or maximum
error is considered \cite{dueck}.  Classically, this separation
between achievable average and maximum error occurs only when we
restrict to deterministic encodings; \mscite{DW05} points out that the
capacity regions for maximum and average error are the same when we
let randomness be introduced into the encodings.  The main result of
our paper can thus be thought of as a coherent version of
\mscite{DW05}.

\section{Implications on the definition of coherent classical
communication}\label{app:implications}

There are two ways to define a cbit.  One is in terms of an abstract
operation $\ket{x}_{\A}\ra\ket{x}_{\B}\ket{x}_{\E}$ for $x\in\{0,1\}$.
Another is more operational, that some sequence of operations $\cP_n$
can send $n$ cbits with error $\epsilon_n \ra 0$ if $\cP_n(\ket{x}_{\A})
\upto{\epsilon_n} \ket{x}_{\B}$, for $x$ an $n$-bit string.
The fact that the operational and abstract definitions are equivalent
allows us to think about classical communication in both ways 
interchangeably.  

Similarly we can define a cobit either as an abstract operation
$\ket{x}_{\A}\ra\ket{x}_{\A}\ket{x}_{\B}$ for $x\in\{0,1\}$, or by
saying that $\cP_n$ can send $n$ cobits with error $\epsilon_n \ra 0$
if $\cP_n$ can send $n$ cbits with error $\epsilon_n$ {\em and}
$\cP_n$ is an isometry.  By Prop~1 of \cite{ccc}, these definitions
are equivalent for one-way communication.  Thm~\ref{thm:bidi-ccc} of
this paper shows that these definitions are now equivalent for two-way
communication.  This justifies the name ``coherent classical
communication''; a cobit really is no more and no less than a cbit
sent through coherent means (i.e. a unitary gate or isometry).

\vspace*{5ex}

{\bf Acknowledgments:}
We are grateful to the Perimeter Institute for their hospitality while
we did this work.  Feedback from the anonymous referees was much
appreciated.  Thanks to Igor Devetak, Andreas Winter, and Jon Yard for
useful discussions, especially on the relation between the worst-case and
the average-case errors and on the significance of \cite{dueck}.  AWH
acknowledges partial support from the NSA and ARDA under ARO contract
DAAD19-01-1-06.  DWL acknowledges support from the Tolman Endowment
Fund, the Croucher Foundation, and the US NSF under grant
no.~EIA-0086038.

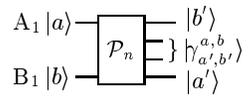
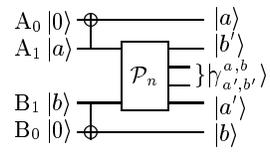
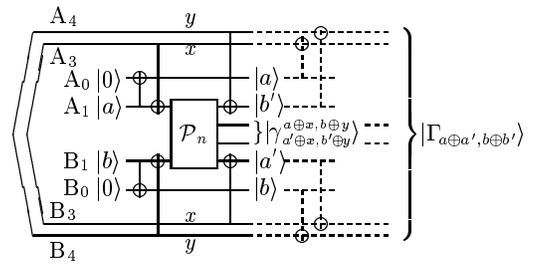
\end{document}